# A Pair-wise Key Distribution Mechanism and Distributed Trust Evaluation Model for Secure Data Aggregation in Mobile Sensor Networks

Natarajan Meghanathan
Associate Professor of Computer Science
Jackson State University, Jackson, MS 39217, USA
E-mail: natarajan.meghanathan@jsums.edu

**Abstract**

We propose a secure data aggregation (SDA) framework for mobile sensor networks whose topology changes dynamically with time. The SDA framework (designed to be resilient to both insider and outsider attacks) comprises of a pair-wise key establishment mechanism run along the edges of a data gathering tree and a distributed trust evaluation model that is tightly integrated with the data aggregation process itself. If an aggregator node already shares a secret key with its child node, the two nodes locally coordinate to refresh and establish a new pair-wise secret key; otherwise, the aggregator node requests the sink to send a seed-secret key message that is used as the basis to establish a new pair-wise secret key. The trust evaluation model uses the two-sided Grubbs' test to identify outlier data in the periodic beacons collected from the child nodes/neighbor nodes. Once the estimated trust score for a neighbor node falls below a threshold, the sensor node "locally" classifies its neighbor node as a "Compromised or Faulty" (CF) node, and discards the data or aggregated data received from the CF node. This way, the erroneous data generated by the CF nodes could be filtered at various levels of the data gathering tree and are prevented from reaching the root node (sink node). Finally, we assess the effectiveness of our trust evaluation model through a comprehensive simulation study.

**Keywords:** Trust Evaluation, Key Establishment, Mobile Sensor Networks, Data Aggregation, Simulations

## 1 Introduction

A wireless sensor network (WSN) comprises of hundreds to thousands of sensor nodes that collect data from the environment for a given task, and one or more sinks (a.k.a. Base Station, BS) responsible for administering and collecting data from the sensor nodes [1]. WSNs are often deployed to detect a parameter or event of common interest (like temperature, fire, intrusion, etc) and the sink needs only one representative data of the entire area being monitored. Hence, instead of requiring the sensor nodes to individually send their data (either directly or through multi-hop paths), it would be more efficient (resource-wise) to gather data from all of these sensor nodes and send only one aggregated version of the data (say the minimum, maximum, sum, etc) to the sink. In this context, several data aggregation protocols (e.g. [2][3][4]) have been proposed for WSNs to eliminate the redundancy in data transmission and thereby reduce the communication and energy overhead at the sensor nodes.

Data aggregation in WSNs is typically conducted using a tree topology, as it is the most energy-efficient in terms of the number of link transmissions; there is only one path from any node to the root leader node (no scope for duplicate packets). Each intermediate non-leaf node in a data gathering tree (DG-tree) acts as an aggregator, fusing the data collected from its immediate child nodes and forwarding the aggregated data to its own upstream parent node in the tree. This way, data is processed and fused at several hops on the way to the leader node, which eventually forwards the aggregated data to the sink.

However, hop-by-hop aggregation of data along the tree is prone to false data injection attacks. Once under his control, an adversary can reprogram a sensor node with malicious code to disrupt the normal functioning of the network. For example, a compromised sensor node could inject one or more spurious data packets that could corrupt the aggregated data that is on its way further up a DA-tree. Though several works in the literature (e.g., [5][6]) have addressed the problem of secure data aggregation, their focus has been on static sensor networks wherein the nodes do not move.

In this research, we focus on secure data aggregation in wireless mobile sensor networks (WMSNs) wherein the sensor nodes move randomly, independent of each other. With the proliferation of electronic devices like Smart Phones and Personal Digital Assistants that are embedded with sensors to measure the temperature, location, humidity and other vital parameters of interest, there has been considerable interest in the sensors community to study data aggregation in the presence of node mobility. Due to the dynamically changing topology, the communication protocols and their secure variants developed for static sensor networks are not directly applicable for WMSNs. Nevertheless, several network disruption attacks (like node capture and false packet injection), packet interception attacks and denial of service attacks (like energy-depletion) are very much possible in WMSNs. In this context, we propose to develop a secure data aggregation framework for WMSNs that comprises of a rigorous trust-evaluation model and pair-wise secret key establishment process for the mobile sensor nodes. Ours will be the first such comprehensive secure data aggregation framework for WMSNs. Existing work, if any exists [7], has dealt with the above two mechanisms only in an isolated fashion, and not together.

The following are the objectives of the proposed secure data aggregation (SDA) framework for WMSNs: (1) Protect from outsider attacks through secure, energy-efficient in-network data aggregation that provides confidentiality, integrity and authentication; (2) Protect from insider attacks through an embedded trust-evaluation model that can be used to detect and mitigate network disruption and energy depletion attacks as well as identify malicious compromised nodes and faulty nodes; and (3) Maximize the number of pair-wise secret keys established between sensor nodes, facilitated through node mobility and communication as part of data aggregation.

The two key characteristic features that are embedded within the proposed SDA framework are: (1) Pair-wise Key Distribution: SDA facilitates establishment of pair-wise secret keys between as many sensor nodes as possible; and to be robust from node capture and packet interception attacks, the pair-wise secret keys need to be refreshed by the concerned nodes every time a new data aggregation tree (DA-tree) is setup. With node mobility [8], we foresee frequent reconfiguration of the DA-trees and hence scope for establishment and renewal of pair-wise secret keys between several node pairs. (2) Trust Evaluation of Sensor Nodes: With the trust-level of the sensor nodes evaluated as part of the data aggregation process itself, the SDA framework is robust to network disruption and denial of service attacks without any additional communication overhead other than those incurred for the secure data aggregation process designed to provide confidentiality, integrity and authentication. Note that in this report, the terms data aggregation and data gathering as well as the acronyms DA-tree and DG-tree are used interchangeably. They mean the same.

The rest of the paper is organized as follows: Section 2 reviews the literature on pair-wise key establishment mechanisms for static and mobile sensor networks and outlines the unique characteristics of our proposed mechanism. Section 3 presents the design of our proposed DA-tree based pair-wise key establishment mechanism for mobile sensor networks. Section 4 presents the distributed trust evaluation model. Section 5 presents a detailed simulation study of the trust evaluation model with respect to stability and minimum-distance spanning tree based data gathering algorithms and various parameters of the model. Section 6 concludes the paper. Section 7 outlines tasks for future work.

## 2  Related Work on Pair-wise Key Establishment and Our Contribution

Most of the literature available for pair-wise key establishment in sensor networks correspond to static sensor networks, where all nodes are static (do not move). We describe below a generic pair-wise key establishment model for static sensor networks, followed by a discussion of selected strategies proposed to improve the generic model.

## 2.1 Generic Pair-wise Key Establishment Model for Static Sensor Networks

The schemes proposed for static sensor networks go through the following two-three phases: (i) a key pre-distribution phase; (ii) direct key establishment phase (shared key discovery), followed by a path-based key establishment phase. In the key-distribution phase (prior to deployment), each node is assigned a fraction of the keys (called key ring) from a larger key pool with the premise that there is a reasonable chance that two neighbor nodes share at least one common key. After deployment, the neighbor nodes exchange their key rings to find common key and share them, if any is found. If two neighbor nodes could not agree on a common key as part of the direct key establishment phase, they establish a path through one or more intermediate nodes such that there is a shared key between two successive nodes on the path. A secret key is then established between the neighbor nodes through such a multi-hop path. The integrity of the nodes constituting the path is critical to the confidentiality of the shared key established between the two neighbor nodes, and may further impact the secret key established in the individual neighborhood of these nodes. Several improvements to the above basic pair-wise key establishment model for static sensor networks have been proposed in the literature and we discuss here some of the representative schemes:

## 2.2 Eschenauer and Gligor Scheme

If the key rings are exchanged (during the key-setup phase) using a simple local broadcast, a casual eavesdropper could identify the key sets of all the nodes in the network and could identify an optimal subset of nodes to compromise in order to discover a larger subset of the key pool. To counter such eavesdropper attacks, Eschenauer and Gligor (EG-scheme) [15] proposed a puzzles-based approach (e.g., the Merkle puzzle [16]) wherein a sensor node could issue a puzzle for each key in its key ring to the neighbor nodes. Any neighbor node that responds with the correct answer for a client puzzle is identified as correctly knowing the associated key.

## 2.3 q-Composite Keys Scheme

To improve network resilience to node capture attacks, Chan et al [17] proposed a q-composite keys scheme, as a modification to the basic EG-scheme, wherein q common keys, with q > 1, are needed to overlap between the key rings of two nodes to setup a secure direct link. The motivation behind this scheme is that as the number of keys (in the key rings) required for overlap increases, it becomes exponentially harder for an attacker with a given key set to break a link. However, the tradeoff is that the number of keys forming the key pool has to be reduced in order to preserve the probability that two nodes share sufficient keys to establish a secure link. This facilitates an attacker to gain a larger sample of the key pool by breaking fewer nodes.

## 2.4 Multi-path Key Reinforcement

If two nodes A and B use a common key *k* in their key rings as the secret key for their communication link, the security of the link could be jeopardized if one or more nodes (other than A or B) that have the secret key *k* in their key rings are captured. To minimize the consequences of such node capture attacks, Anderson et al [18] propose a multi-path key reinforcement scheme wherein node A discovers multiple node-disjoint paths to its neighbor node B and sends a random sequence of bits (of the same length as the original secret key *k*), one sequence per node-disjoint path. If $v_1, v_2, \ldots, v_j$ are the random bit sequences sent along *j* node-disjoint paths from A to B, then the new secret key *k'* computed and used by both A and B for communication across their link would be a simple XOR of the original secret *k* with all the random bit sequences $v_1, v_2, \ldots, v_j$. That is, $k' = k \oplus v_1 \oplus v_2 \oplus \ldots \oplus v_j$. Unless otherwise, an attacker eavesdrops on all the *j* node-disjoint paths, it would be difficult to extract *k'* from *k*. However, the longer a node-disjoint path, the more vulnerable the path to reveal a random bit sequence (if the attacker manages to capture at least one intermediate node on a path, he can get access to the random bit sequence sent along that path). Hence, Chan et al [17] suggest the use of a 2-hop multi-path scheme for deriving a new secret key from a shared secret key in the key rings. The route discovery overhead associated with setting up multi-hop node-disjoint paths would also be minimal in the case of a 2-hop multi-path scheme: the two end nodes can exchange their neighbor lists (in the underlying communication network topology) and choose to route the random bit sequences through those neighbor nodes with which they share a secret key in their key rings.

### 2.5 Issues with Pair-wise Key Establishment in Mobile Sensor Networks

There are not many schemes proposed for key distribution in mobile sensor networks. The topology of the mobile sensor networks changes dynamically with time. Due to node mobility, there are good chances that any two nodes could be neighbors at some point of time. As a result, we may have to run path-based key establishment at any time (and not just after the initial node deployment). For the path-based key establishment phase to be successful, we need connectivity of the underlying network of direct links. It is very difficult to select a proper size for the key rings and the global key pool so that there can be at least one common key between an appreciable number of neighbor nodes to establish the direct links between nodes (as part of the direct key establishment phase) leading to connectivity of the underlying network. In the literature on mobile sensor networks, we came across a location-based strategy [19] that requires sensor nodes to be aware of their location (post-deployment) and use it as part of the pair-wise key establishment process. However, it is too energy-draining for a sensor node to use GPS-kind of mechanisms to regularly find about its location.

### 2.6 Unique Characteristics of our Proposed Pair-wise Key Establishment Scheme

There is no need to generate a key ring for every sensor node. Instead, we primarily assume every sensor node to share a secret key with the sink (see section 3.1 for a detailed list of assumptions). Our proposed scheme does not require a sensor node to establish direct link or path-based keys with each of its neighbor nodes. Instead, we first setup a data aggregation tree (DA-tree) of the underlying sensor network and require an aggregator node (intermediate node) of the tree to take up the responsibility of establishing pair-wise secret keys with its child nodes, if none exists so far, in co-ordination with the BS (sink). If an aggregator node already shares a secret key with a child node, it establishes a new secret key using the current shared secret key as the basis. By using the DA-tree as the underlying communication topology for pair-wise key establishment, we avoid the communication overhead as part of path-based key establishment in the neighborhood of every sensor node (associated with the pair-wise key schemes for static sensor networks). The communication between the aggregator node and the sink is facilitated through the paths that are part of the DA-tree itself and there is no need to establish separate paths between the sink and each of the aggregator nodes for pair-wise key establishment with their child nodes. Moreover, unlike the earlier mechanism proposed in the literature (discussed in Section 2.5), our proposed pair-wise key establishment mechanism does not require the sensor nodes to be location-aware.

## 3 Pair-wise Key Establishment along a Data Aggregation Tree

### 3.1 Key Assumptions

The key assumptions are: (1) The sink BS is secure and it shares a 128-bit secret key with each sensor node. (2) The BS broadcasts any information to one or more sensor nodes in a secure fashion through the well-known µTESLA [9] protocol that also provides authentication. (3) Each node stores a Key Cache in its memory. The Key Cache of a node consists of the 128-bit pair-wise secret keys the node shares with one or more of the other nodes in the network and the sink. (4) The DA-tree is securely formed each time when needed. There is no scope for attack or impersonation during its construction.

### 3.2 Sequence of Steps in Pair-wise Key Establishment

**1.** *Construction of the DA-Tree*: The sink initiates the construction of the DA-tree by sending a trigger signal to the leader node (the node with the largest available energy) to broadcast a Tree-Construction message to its neighborhood. When a sensor node receives the Tree-Construction message for the first time, it sets the sender of the message as its predecessor upstream node in the tree and broadcasts the message further to its neighbors, and discards any further Tree-Construction messages received. This way, the Tree-Construction message is broadcast by every sensor node (but exactly once) so that it can propagate throughout the network to eventually establish a spanning topology tree rooted at the leader node. The child nodes send a Parent-Update message to inform their parent node that they have chosen the latter as their immediate upstream node in the DA-tree.

**2.** *Maintenance of the DA-Tree*: Each intermediate node of the DA-tree monitors the links to its downstream nodes (expecting periodic beacons from the neighbor nodes). When a downstream link breaks, the upstream node sends a secure Tree-Error message (encrypted and decrypted every hop on the path with the pair-wise secret keys that would have been established by then for the DA-tree using steps 3 through 5) to the leader node, which forwards the message further to the sink. The sink initiates another tree construction process (Step 1).

**3.** *DA-Notification Message from the Aggregator Node to the Sink*: If the aggregator node of the DA-tree does not share a secret key with one or more of the child nodes, then the aggregator node sends a DA-Notification message (encrypted with the shared secret key) to the BS, containing the IDs of the child nodes with which it does not yet share a secret key. The DA-Notification message also contains a Nonce (randomly generated by the aggregator node) that needs to be incremented and included in the Seed-Secret-Key message sent as a response by the sink BS.

**4.** *Seed-Secret-Key Message from the Sink to the Aggregator Node*: The sink BS responds to the DA-Notification message with a Seed-Secret-Key message that contains different components: one for the aggregator node and one for every requested child node (with which the aggregator node needs to establish a new pair-wise secret key). The Seed-Secret-Key message component meant for the aggregator node – contains 64-bit random numbers generated by the BS (one for each of the aggregator node and its requested child nodes) as well as the incremented value of the Nonce sent in the DA-Notification message. The Seed-Secret-Key message components meant for the individual child nodes include the 64-bit random number generated for the aggregator node as well as the aggregator node ID and a 64-bit random number generated for the individual child node. The Seed-Secret-Key message component meant for a node is encrypted using the secret key shared by the sink BS with the individual node.

**5.** *Agreement on a New Pair-wise Secret Key*: Upon the receipt of the Seed-Secret-Key message from the sink BS, an aggregator node forwards to the appropriate child nodes – their respective encrypted message components as it is. A child node decrements the Seed-Secret-Key component message (addressed for it) using the secret key shared with the BS and ensures that the message component includes the ID of the aggregator node that forwarded the Seed-Secret-Key message. The child node then generates a 128-bit pair-wise secret key (generated on the fly by the child node) as well as a Nonce value (meant for the aggregator node to later send an incremented Nonce value in its response and get authenticated for receiving the new pair-wise secret key) and encrypts them using a temporary key, which is basically a product of the random numbers (generated by the sink BS and included in the Seed-Secret-Key message component) for the individual child node and the aggregator node. The child node sends this encrypted pair-wise secret key to the aggregator node as part of a New Pair-wise Key Establishment message. Upon receiving this message from a child node, the aggregator node computes on its own the temporary key (product of the random number sent to it by the sink BS and the random number meant for the appropriate child node) and decrements the message to extract the new pair-wise secret key and the Nonce value. The aggregator node responds with the New Pair-wise Key Acknowledgment message (encrypted with the new pair-wise secret key) containing the incremented Nonce value. After validating the incremented Nonce value in the acknowledgment message, the child node is convinced of the establishment of the pair-wise secret key with the aggregator node.

**6. *Refreshment of a Pair-wise Secret Key*:** If an aggregator node already shares a pair-wise secret key with a child node established earlier due to their association in some data gathering tree (DG-tree) of the past, the aggregator node refreshes this association by establishing a new pair-wise secret key using the currently shared pair-wise secret key as the basis. In this pursuit, the aggregator node sends the child node a Pair-wise Key Refresh Request message that contains a randomly generated Nonce encrypted using the currently shared pair-wise secret key with the child node. The child node decrypts the message using the shared pair-wise secret key, and sends the following as part of a Pair-wise Key Refresh Response message: the incremented value of the Nonce received from the aggregator node, a new 128-bit pair-wise secret key (generated on the fly by the child node) as well as a new Nonce value (to be used to validate the aggregator node's receipt of the new pair-wise secret key) – all of these encrypted with the most recently shared pair-wise secret key. The aggregator node decrypts the Pair-wise Key Refresh Response message to extract the new pair-wise secret key and confirms the receipt of the same by sending back (to the child node) an incremented value of the Nonce (generated and included by the child node in the Refresh Response message) encrypted with the new pair-wise secret key sent as part of a Pair-wise Key Refresh Acknowledgment message.

### 3.3 Example for Pair-wise Key Establishment

We now illustrate below the contents of the DA-notification message, seed-secret-key message and the sequence of messages that will be exchanged to establish or refresh the pair-wise secret key between an aggregator node and its child nodes. Note that the DA-notification message contains only the IDs of those child nodes with which the aggregator node does not already share a pair-wise secret key. For the sake of illustration, we refer to the aggregator node as A and its child nodes as B, C and D. We also assume that node A already shares a pair-wise secret key with node B and the key is to be refreshed (Figure 2). Node A requests the BS for a seed-secret-key only to establish pair-wise secret keys with C and D (Figure 1 shows the message exchange between A and D to establish a pair-wise secret key).

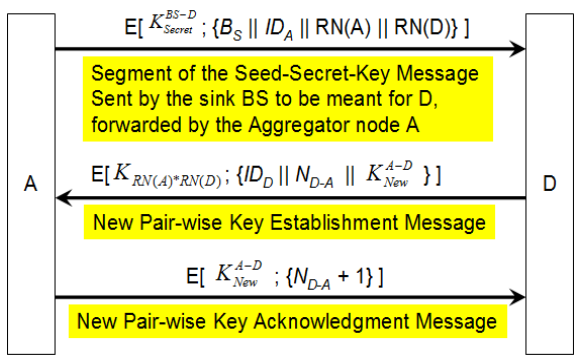
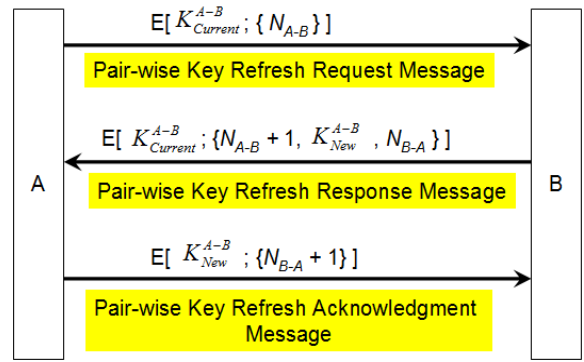

**Figure 1:** Message Exchange: Aggregator – Child Nodes to Establish a New Pair-wise Secret Key

**Figure 2:** Message Exchange: Aggregator - Child Nodes to Refresh/Renew a Pair-wise Secret Key

*DA-Notification Message*:
E[ $K_{Secret}^{BS,A}$ ; {$ID_A$ || Nonce ($N_A$) || Children($ID_C$, $ID_D$) } ]

*Seed-Secret-Key Message*:
E[ $K_{Secret}^{BS,A}$ ; {*BS* ‖ Nonce ($N_A$ + 1) ‖ RN(A) ‖ RN(C) ‖ RN(D)}] ‖ E[ $K_{Secret}^{BS,C}$ ; {*BS* ‖ *ID_A* ‖ RN(A) ‖ RN(C)} ]
‖ E[ $K_{Secret}^{BS,D}$ ; {*BS* ‖ *ID_A* ‖ RN(A) ‖ RN(D)} ]

## 4  Distributed Trust Evaluation Model

We consider spanning tree-based data aggregation for mobile sensor networks. The data gathering tree (DG-tree) is formed using a spanning tree of the network that also includes the sink node as one of the constituent nodes. We run the Breadth First Search algorithm [10] on the spanning tree, starting from the sink node as the root node (leader node). As the edges of the spanning tree are traversed from the root node, the directions get assigned to the edges: leading to the nodes being classified as intermediate nodes and leaf nodes.

Data aggregation starts from the leaf nodes. The nodes in the DG-tree are classified according to their position in the tree. The height of the DG-tree is the distance from the sink node to the leaf node that is located at the farthest distance (measured in terms of the number of edges). The root node is considered to be at level 0 and the immediate downstream child nodes of the root node are said to be at level 1, and so on. A leaf node forwards its individual beacon data (also considered to be the aggregated data) to its immediate upstream parent node. For an intermediate node to aggregate data, it should have received the aggregated data from all of its immediate downstream child nodes. The aggregated data (in the case of a child node that is a leaf node, the aggregated data is the beacon data) at an intermediate node is the sum of the aggregated data received from its immediate downstream child nodes and its own beacon data; this aggregated data is forwarded by the intermediate node to its own upstream parent node in the DG-tree.

As the sensor nodes could become compromised (insider attacks) or faulty and generate erroneous data, it is imperative to weed out such corrupt data and facilitate the data aggregation process to bypass the compromised or faulty nodes (referred to as CF nodes). We propose a trust evaluation model that is based on assigning raw trust scores to sensor nodes based on the validity of the data generated by the nodes themselves (test for outliers). We assume a CF node to randomly generate data from a broader range and that is likely to be different from the range for normal data that is expected of the node. A trust score is assigned for a node based on the outliers detected in its data sequence and the trust score is dynamically updated for every round of data aggregation. A node whose sequence of data has more outliers, compared to the regular data, is likely to be untrustworthy, and is flagged a CF node after its trust score falls below a threshold. As part of our research, we have also analyzed the impact of different operating parameters of the Secure Data Aggregation (SDA) framework on the effectiveness of the trust evaluation model, measured with respect to the number of rounds (median value) incurred to detect the presence of CF nodes as well as the average value of the aggregated data at the sink (wherein data aggregation occurs in the presence of the CF nodes).

An intermediate node considers the aggregated data received from a downstream child node in its data aggregation calculations only if the downstream child node has not been classified as a CF node. Aggregated data received from a downstream child node that is classified as a CF node is not considered for aggregation. To speed up the execution of the whole SDA framework, once a node is classified as a CF node, even the periodic beacon data received from that node is dropped and the trust value for the CF child/neighbor node is not calculated.

To keep track of the number of individual nodes that have contributed to the aggregated data received from a downstream child node, we include a *numSDAUsedNodes* field in the header of the aggregated data packet. The value of the *numSDAUsedNodes* field in the header of the aggregated data packet is the sum of the values of the *numSDAUsedNodes* fields in the aggregated data packets received from the immediate downstream child nodes plus 1 (the '1' corresponds to the intermediate parent node that is aggregating the downstream child node data with its own beacon data). The *numSDAUsedNodes* field value for a leaf node is 1.

## 4.1 Example for Data Aggregation in the Presence of CF Nodes

We exemplify the data aggregation process and the treatment meted out to aggregated data received from the CF nodes in Figure 3. There are a total of 12 nodes in the network, including the root node of the DG-tree, which is the sink node (shown in green color). The regular nodes (non-CF nodes) are shown in light blue color and the CF nodes are shown in orange color respectively. The numbers outside the nodes/circles indicate the raw sensed data at the individual sensor nodes and the *numSDAUsedNodes* values in the header of the aggregated data packet at the level corresponding to the node. All leaf nodes have a *numSDAUsedNodes* value of 1. Beacon data or aggregated data from nodes classified as CF nodes are ignored. This could be observed in the calculations done at the intermediate node to the left of the DG-tree in Figure 1. The beacon data (of value 45) collected from a child CF node is ignored while calculating the aggregated data value (82 + 75 + 90 + 87 = 334) at the intermediate node, including the latter's own beacon data (of value 87). The *numSDAUsedNodes* value is 3 + 1 = 4 (where '3' corresponds to the sum of the *numSDAUsedNodes* values at the child nodes, which are all leaf nodes; and '1' corresponds to the intermediate node itself).

The right part of the DG-tree has no CF nodes; so, the data aggregated at the right child node of the sink node is 93 + 73 + 95 = 261 and the *numSDAUsedNodes* value is 2 + 1 = 3. Note that the sink node rejects the aggregated data value received from the middle child node (second child node, which had generated a data value of 140) that had been already classified as a CF node. The sink node aggregates data from the left and right child nodes (ignoring the aggregated data received from the middle child node) and uses the corresponding *numSDAUsedNodes* to determine the overall average value for the data sensed from the network field. In the above example, the aggregated data value at the sink node is 334 + 261 = 595 and the total value for *numSDAUsedNodes* is 4 + 3 = 7, leading to the overall average value for the data sensed from the network field to be computed as 595/7 = 85.0.

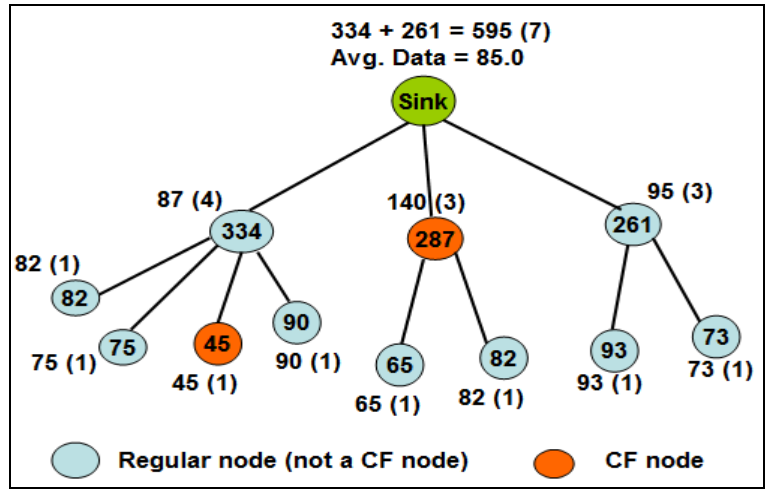

**Figure 3:** Example to Illustrate Data Aggregation in the Presence of CF Nodes

## 4.2 Computation of the Raw Trust Score

An intermediate parent node conducts the trust calculations on each of its immediate downstream child nodes for every beacon data received and updates an estimated average trust score (*Est. avg. trust score*) based on the raw trust score evaluated on the result of the *TrustComputation* procedure run on the newly (latest) added beacon data (referred to as *insertedData*) to the *BeaconWindow*. The raw trust score for the *insertedData* is computed using the Grubbs' two-sided test [11].

The outlier test for the *insertedData* is conducted as follows: We determine the minimum and maximum data (*MinData* and *MaxData*) in the BeaconWindow as well as the *Mean* and standard deviation (*SD*) of the values. We then determine the appropriate t-score for the sample size corresponding to the size of the BeaconWindow. Table 1 lists the t-score reference values [11] used in our outlier detection calculations; the confidence interval is 95%. For sample size (size of the Beacon Window) that is < 5000 and not in Table 1, the appropriate t-score value is obtained through linear interpolation.

**Table 1:** t-Score Table used for Two-Sided Grubbs' Test [11]: 95% Confidence Interval

| # samples | 1 | 2 | 3 | 4 | 5 | 6 | 7 | 8 |
|---|---|---|---|---|---|---|---|---|
| t-score | 12.706 | 4.303 | 3.182 | 2.776 | 2.571 | 2.447 | 2.365 | 2.306 |

| # samples | 9 | 10 | 15 | 20 | 25 | 30 | 40 | 60 |
|---|---|---|---|---|---|---|---|---|
| t-score | 2.262 | 2.228 | 2.131 | 2.086 | 2.060 | 2.042 | 2.021 | 2.000 |

| # samples | 120 | ≥ 5000 |
|---|---|---|
| t-score | 1.980 | 1.960 |

The threshold value ($G_{thresh}$) for the Grubbs' test is calculated as follows:

$$G_{thresh} = \frac{|BeaconWindow|-1}{\sqrt{|BeaconWindow|}} * \sqrt{\frac{tscore*tscore}{|BeaconWindow|-2+(tscore*tscore)}}$$

The *insertedData* is an outlier if it is equal to either the *MinData* or the *MaxData* and satisfies one of the following two inequalities, as appropriate.

$$\frac{|Mean - MinData|}{SD} > G_{thresh} \quad \text{or} \quad \frac{|Mean - MaxData|}{SD} > G_{thresh}$$

If the *insertedData* is classified as an outlier, the *TrustComputation* procedure returns 0; otherwise, returns 1. An intermediate parent node keeps track of the trust scores (stored in a *TrustScoreBuffer*) computed on an intermediate downstream child node in the recent past and during the current association. The size of the *TrustScoreBuffer* is limited by the *MaxTrustScoreBufferSize* variable, an input parameter in our simulations.

**4.3 Computation of the Estimated Average Trust Score**

The calculations to update the *Est. Avg. Trust Score* are conducted every time a raw trust score is returned from the *TrustComputation* procedure and the number of raw trust scores accumulated in the *TrustScoreBuffer* is at least half the *MaxTrustScoreBufferSize*. The *Est. Avg. Trust Score* (maintained at an intermediate parent node for an immediate downstream child node) is the weighted average trust score of two averages of the raw trust scores (0s and 1s) collected in the *TrustScoreBuffer*: one average is computed on all the raw trust score data accumulated due to previous association(s) between the two nodes on DG-tree(s) that had existed before the current DG-tree; the second average is computed on all the raw trust-score data accumulated due to the association between the two nodes on the current DG-tree. The weighted average is computed using a parameter *historyWeight*, whose values range from 0 to 1. The formulation is as shown below:

*Est. Avg. Trust Score*
 = *historyWeight* \*Average (data from TrustScoreBuffer based on prev. associations) +

(1 - *historyWeight*) * Average (data from TrustScoreBuffer based on curr. association)

If the *Est. Avg. Trust Score* is below the *TrustThreshold*, then the sensor node is classified as a CF node. We record the round number it took to identify a node as CF node and keep track of the difference in the number of rounds since the node was set to be a CF node. A CF node is used primarily to determine/establish a data gathering tree, and not for data aggregation. The aggregated data/beacon data received from a CF node are not processed by its parent/immediate upstream node in the DG-tree.

## 5 Simulations

Simulations are conducted in a discrete-event simulator developed by us in Java. The same simulator has been earlier used by us for research [12] in mobile sensor networks. We conduct simulations on a network topology of dimensions 100m x 100m. There are a total of 100 nodes in the network, each operating at a fixed transmission range. The sink is located at one end of the network (at [100, 100]) and is also the root of the data gathering tree (DG-tree). The DG-tree used for data aggregation is based on the MST minimum distance-based spanning tree (MST) or the LET link expiration time-based spanning tree (LET). We use the distributed algorithms proposed in our earlier work [12] to determine the MST and LET-based DG-trees for mobile sensor networks.

Node mobility is according to the Random Waypoint model [13] according to which nodes move randomly, independent of each other: To start with, nodes are uniform-randomly distributed throughout the network topology; each node moves to a randomly chosen target location at a velocity uniform-randomly selected from the range [0...$v_{max}$] and the node moves to the target location with the chosen velocity; after moving to the target location, the node chooses another value for the velocity from the range [0... $v_{max}$] and moves to a new randomly chosen target location. Each node continues to move like this, independent of each other, until the end of the simulation time, which is 1000 seconds. The values for $v_{max}$ used are 3 m/s (low mobility) and 10 m/s (high mobility). For each combination of $v_{max}$, number of nodes and sink location, we generate mobility profiles for nodes offline and input them to the data gathering algorithm under execution.

### 5.1 Simulation Parameters and their Values

The following are the input parameters for the simulation:
- *TransRange* - the fixed transmission range for every sensor node
- *MeanData* - the mean value of data generated by every sensor node
- *STDData* - the standard deviation of data generated by every sensor node
- *MaxBeaconWindowSize* - the maximum size of the beacon window (BW) maintained to store the beacons received from each node in the neighborhood
- *MaxTrustScoreBufferSize* - the maximum size for the Trust Score Buffer (TSB) maintained at an intermediate parent node for every immediate downstream child node. The TSB is used to store the raw trust score values (0s and 1s) calculated based on the beacon data received from the neighbor child nodes at the beginning of each round.
- *TrustThreshold* - the threshold value for the estimated average trust score below which the sensor node is classified as CF node.
- *CFProb* - the probability with which a non-CF node could become compromised or faulty at any round. Once a node becomes a CF node, it continues to remain so till the end of the simulation.
- *MaxCFNodes* - the maximum number of nodes that could become CF nodes
- *historyWeight* - a weight parameter (ranging from 0 to 1) to capture the tradeoff with respect to giving importance to the trust evaluation results collected on a child node at a parent node (in the Trust Score Buffer) during their previous associations vs. the trust evaluations during their current association.

The maximum BW size (Beacon Window size) at each sensor node is set to be 10 and 50. The maximum size of the Trust Buffer (TSB) maintained by a sensor node for each of the other nodes (updated for the child nodes) is set to be 10, 30 and 50. The *Check for CF status* is conducted for every round of data aggregation, once the TSB Size reaches half of its maximum value (set to one of the above three values). The value for the *TrustThreshold* (the minimum value for the estimated average trust score below which the node is classified as CF node) is set to values of 0.5, 0.7 and 0.9. The *historyWeight* parameter is set to values of 0.3, 0.5, 0.7, 0.9 and 1.0: If the *historyWeight* parameter is set to 1.0, it implies that trust score values determined by a parent node for a child node during their current association would not be considered and the Check for CF Status will be conducted only based on the trust scores estimated during the previous association(s) between the two nodes. On the other hand, a *historyWeight* value of 0.3 implies that relatively much larger importance (100 - 30% = 70% weight) will be given to the trust scores determined during the current association between the two nodes. The number of CF nodes is set to values of 20 and 40; as the total number of nodes is 100, this corresponds to operating the network with 20% and 40% of nodes as CF nodes respectively. Until we have turned on the above said number of nodes as CF nodes, any non-CF node in the network has an equal chance of becoming a CF node; the probability (*CFProb*) with which any non-CF node can be come a CF node is set to 0.005. Simulations are considered for fixed transmission range per node values of 25 and 35. Though the TrustScoreBuffer is filled with the raw trust score data computed on the individual data in the BeaconWindow, the parameters MaxTrustBufferSize and MaxBeaconWindowSize are treated independently and the values are assigned individually without any dependency on each other.

Each data point in Figures 2 through 9 is average of results obtained from 100 mobility profiles. For each mobility profile, we run the simulation for 1000 seconds. A simulation (corresponding to either the MST or LET-based DG trees) is run for fixed values of *BW Size* (10 and 50), *TSB Size* (10, 30 and 50), *TrustThreshold* (0.5, 0.7 and 0.9), *historyWeight* (0.3, 0.5, 0.7, 0.9 and 1.0), transmission range (25m and 35m), maximum node velocity ($v_{max}$ = 3 m/s and 10 m/s) and number of CF nodes (20 and 40), resulting in a total of 2*3*3*5*2*2*2 = 720 combinations of scenarios for which the results are shown in Figures 2 through 9. We have also collected results for *TrustThreshold* values of 0.6 and 0.8 and *TSB Size* of 70. We do not present the results obtained for these combinations of *TrustThreshold* and *TSB Size* values due to lack of space and the observation of trends similar to those combinations chosen for presentation.

### 5.2 Data Generation Model

The data measured from the network field is assumed to be the temperature of the field. We want only one representative data - either the maximum, minimum or the mean temperature of the network field. In the simulations, we assume the data is sensed at each node and broadcast to its neighbor nodes (within the transmission range) through beacons. If a node is not a CF node (that is a benevolent regular node, neither compromised nor faulty), then the temperature data generated by the node is between the range [*MeanData* - *STDData* ..... *MeanData* + *STDData*]. If a node is a CF node, then the data generated by the node is within the range from [0 .... 5**MeanData*].

For the sake of the simulations, data at a CF-node is generated for sensing by each node according to a uniform distribution with mean 80 and standard deviation 20. Accordingly, for every data point, we generate a random number from the range $x \in [0 ... 1]$ and the individual data point generated is then 80 ± 20*x*. Data at a CF node is generated uniform-randomly from the range [0... 5*80].

### 5.3 Selection of Compromised or Faulty (CF) Nodes

The simulations are conducted with 100 nodes (*numNodes*). Among these 100 nodes, a fixed number of nodes (*MaxCFNodes*) are set to become CF nodes (Compromised or Faulty) nodes at some time instant during the simulation and will continue to remain as CF nodes since then. Any node could become a CF node. Until the *MaxCFNodes* number of nodes have become CF nodes, we run the *CFEnable* routine at the beginning of each round, starting from round # 10 (we wait for at least 10 rounds before setting any node to become a CF node). The probability (*CFProb*) with which a node can become a CF node is an input parameter to our simulation. Until the *MaxCFNodes* number of nodes are compromised or become faulty, any node can become a CF node with a probability of *CFProb* when the *CFEnable* routine is run.

The CFEnable routine runs as follows: For every node (node id 0 to *numNodes* - 1) that is not yet compromised or faulty, we generate a random number from 0 to 1. If the random number generated is less than or equal to *CFProb*, then that node is considered to have become a CF node and is added to the list of CF nodes. Any number of nodes (0 or more) could become a CF node during a particular round.

### 5.4 Data Gathering per Round

For each mobility profile, we run the simulation for 1000 seconds. There are 4 rounds of data gathering per second, set to be at a fixed rate: the time between two successive rounds of data gathering is 0.25 seconds. If a DG-tree existed during the previous round, we check whether all the edges in the DG-tree exists during the current round. An edge is said to exist if the Euclidean distance between the two constituent end nodes of the edge is less than or equal to the transmission range of the nodes. If the DG-tree exists during a round, then we are ready for data aggregation. We let each sensor node to generate the data, whose values depend on whether the node is a CF node or not. A node is assumed to broadcast the generated data to its neighborhood; and thereby, a parent node receives a beacon from each of its child neighbor nodes (i.e., the immediate downstream nodes). If the DG-tree that existed during the previous round(s) no longer exists, then we determine the appropriate DG-tree (using either the MST-based or LET-based DG-tree algorithm of [12]) and then initiate data aggregation for the current round.

### 5.5 Beacon Management

The newly received beacon from a child node (of a unique node ID) is stored in the *BeaconWindow* (BW) maintained for the particular child downstream node. Before inclusion of the new beacon data, if BW Size equals the *MaxBeaconWindowSize*, then the earliest recorded data in the BW is purged and the newly received beacon data is appended to the BW. If the BW Size is less than *MaxBeaconWindowSize*, then the newly received beacon is simply appended to the BW. The raw sensed data collected from a child neighbor node and accumulated in the BW maintained for the node is used to assess the trust value of the child node.

### 5.6 Performance Metrics

We measure the following for each simulation scenario: (1) *The median value for the number of rounds* to detect the CF nodes (for every CF node correctly identified, we record the difference between the round at which the node is identified to be a CF node and the round at which the node was set as a CF node, and determine the median value of all these differences), averaged over the 100 mobility profiles. (2) *Average of the data values collected at the sink node* based on the data aggregated from the non-CF nodes and the number of non-CF nodes that participated in the data aggregation process.

### 5.7 Simulation Results and their Interpretation

The MST-based DG-trees have been observed to be more energy-efficient [14]; but more unstable in the presence of node mobility [12]. The LET-spanning tree based DG-trees are inherently more stable due to the presence of links that are predicted to have a relatively larger lifetime [12].

For a given simulation condition, in the presence of CF nodes, the LET-based DG-trees have been observed to yield more accurate values for the total aggregated data compared to those determined using the MST-based DG-trees. We also observe the LET-based DG trees to incur a lower median value for the number of rounds to detect the CF nodes (both for 20% and 40% CF node scenarios). For a fixed $v_{max}$, % CF nodes, transmission range and TSB Size, there is no significant impact of the Beacon Window Size

(i.e., for BW Size values of 10 or 50) on the median number of rounds to detect the CF nodes as well as on the average value for the data aggregated at the sink in the presence of these CF nodes.

With regards to the impact of the *historyWeight* parameter, we observe that the SDA Framework incurs the lowest value for the median number of rounds (to detect CF nodes) when operated with a *historyWeight* parameter value of 0.3. The effect of the *historyWeight* parameter is observed more prominently when operated under lower *TrustThreshold* values (especially for 0.5). This is because, when we operate at lower *TrustThreshold* values, it is more likely to take a longer time (# rounds) for the estimated average trust score to fall below the smaller *TrustThreshold* value; hence, it is critical to quickly recognize the erroneous data generated at the CF nodes and detect their presence in the network so that the aggregated data (if the CF node is an intermediate node) / individual data (if the CF node is a leaf node) received from these nodes can be discarded.

With regards to the impact of the *TrustThreshold* and *TSB Size* parameters, for low node mobility scenarios, we observe that larger the *TrustThreshold*, the easier it is to identify and declare a compromised or faulty node as CF node. For smaller values of the *TrustThreshold* parameter, we incur more rounds to detect the presence of CF nodes, as more samples need to be collected when the node is indeed a CF node, especially for high values of the *TSB Size*. Note that the Check for CF Status is done only when the number of raw trust score values stored in the TSB is at least half of the maximum *TSB Size*. Until then, the presence of CF nodes only corrupts the values for the aggregated data. Hence, it is more sensible to operate the network at larger *TrustThreshold* values and lower *TSB Size*.

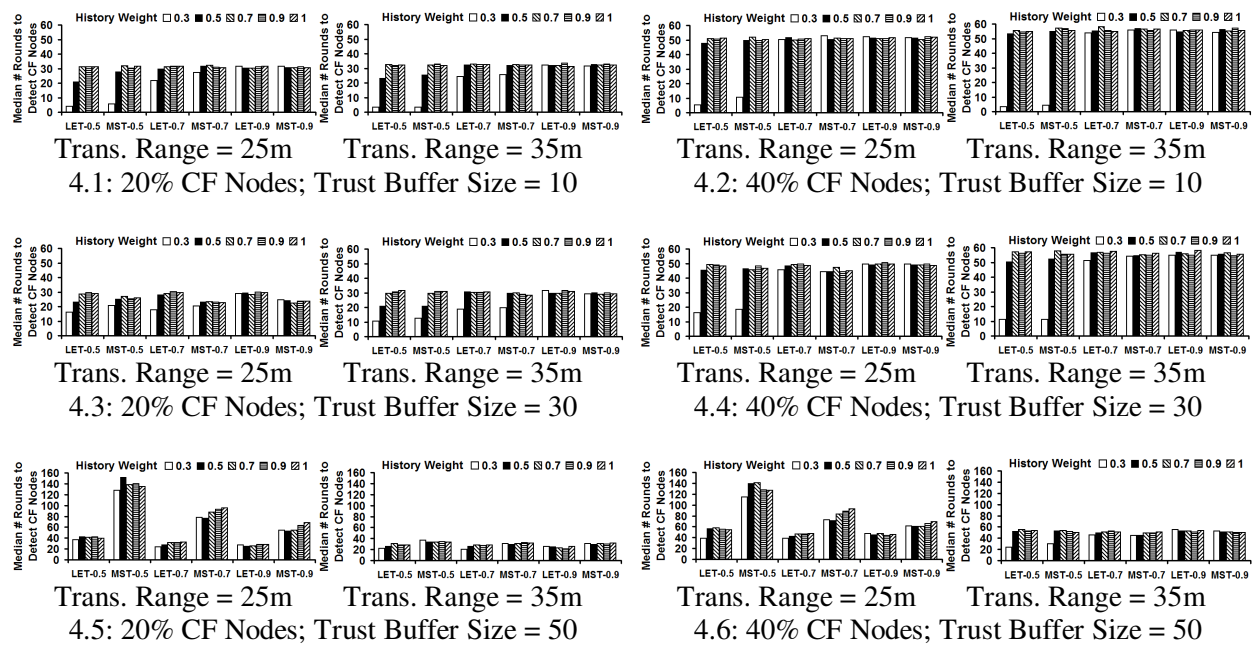

**Figure 4:** Median # Rounds to Detect CF Nodes ($v_{max}$ = 3 m/s, Beacon Window Size = 10)

For low node mobility scenarios, we observe the median # rounds to detect CF nodes does not depend much on the values for the *TrustThreshold* and gets slightly lower for larger values of *TSB Size*. As the DG-trees are likely to exist for a longer time in low node mobility scenarios, the trust estimates are more accurately captured and the CF nodes are more easily identified. There is no need to go through several rounds of data gathering and several refreshes to the TSB before identifying the CF nodes. In other words, there is no significant impact of the *TrustThreshold* and the *TSB Size* on the median number of rounds to detect the CF nodes and the detection is done much earlier in low node mobility scenarios vis-a-vis compared to the time incurred in case of high node mobility scenarios (see analysis below). As a result, there is also less corruption in the data at low node mobility scenarios.

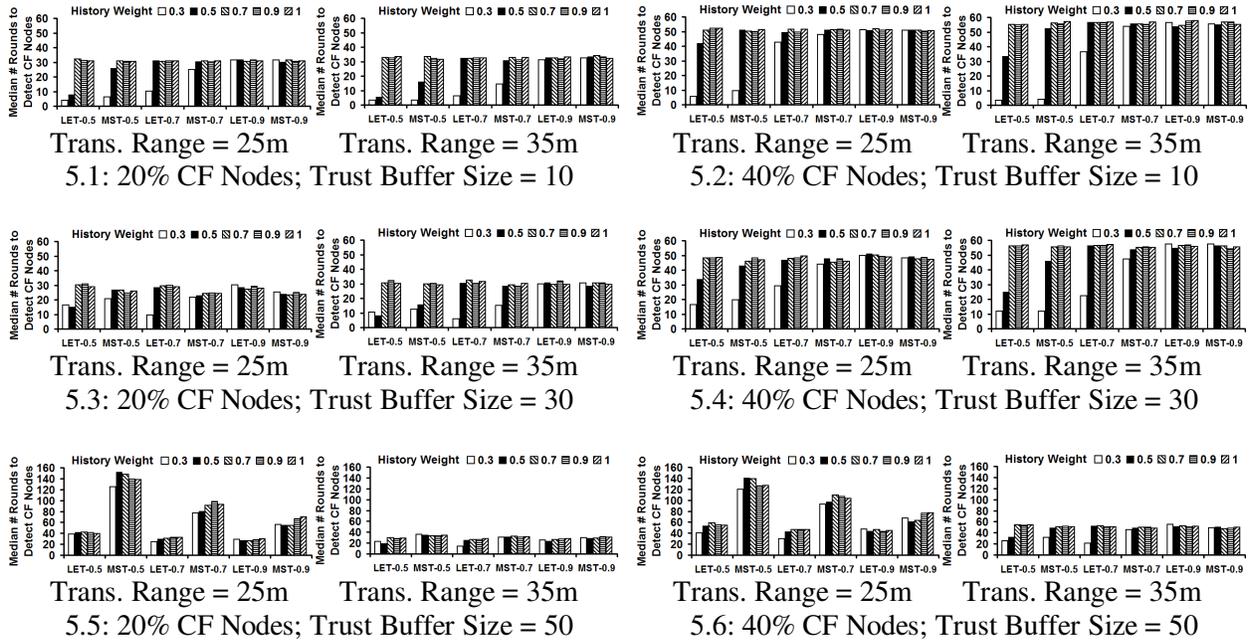

**Figure 5:** Median # Rounds to Detect CF Nodes ($v_{max}$ = 3 m/s, Beacon Window Size = 50)

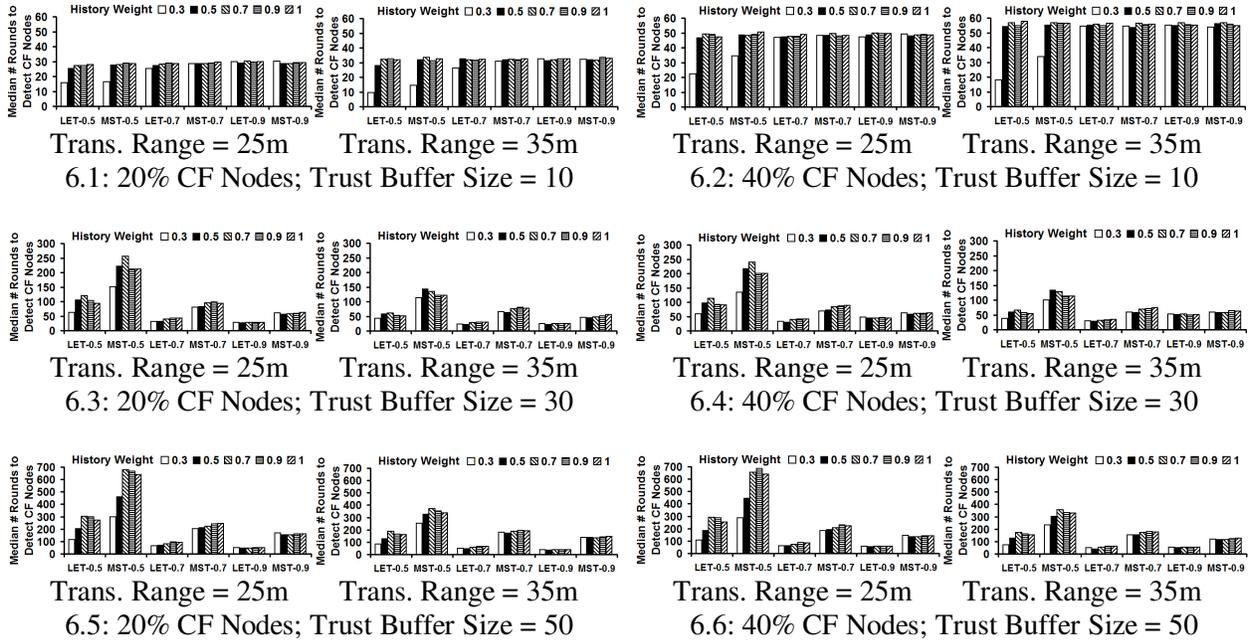

**Figure 6:** Median # Rounds to Detect CF Nodes ($v_{max}$ = 10 m/s, Beacon Window Size = 10)

For larger values of node velocity, the DG-trees are more unstable. As a result, the parent-child association between any two nodes does not last for a longer time. Since the trust calculations are conducted only after the trust buffer has accumulated a reasonable number of trust estimate values

(depending the Trust Buffer Size), the median number of rounds to detect the presence of CF nodes shoots up to significantly high values in the presence of node mobility, especially for the MST-DG trees (could be as large as 6-7 times the values obtained for lower node mobility). The LET-DG trees are more stable and hence incur relatively lower values for the median number of rounds to detect the CF nodes. The longer it takes to detect the presence of CF nodes, the more the corruption to the aggregated data value at the sink.

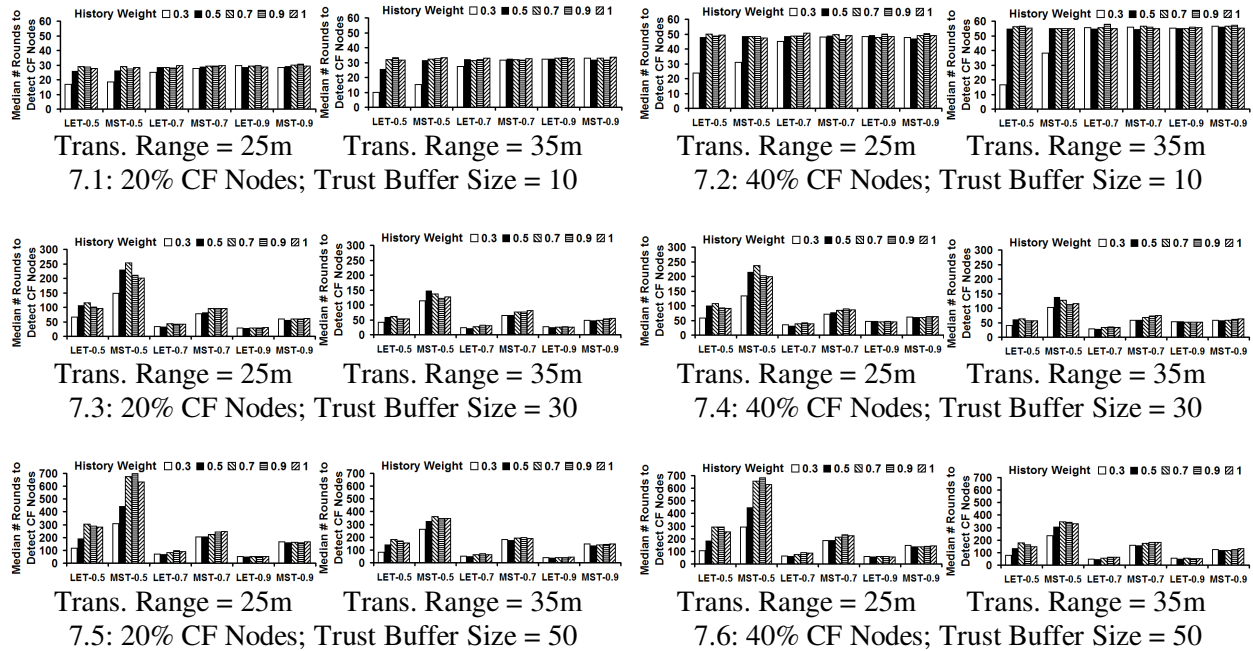

**Figure 7:** Median # Rounds to Detect CF Nodes ($v_{max}$ = 10 m/s, Beacon Window Size = 50)

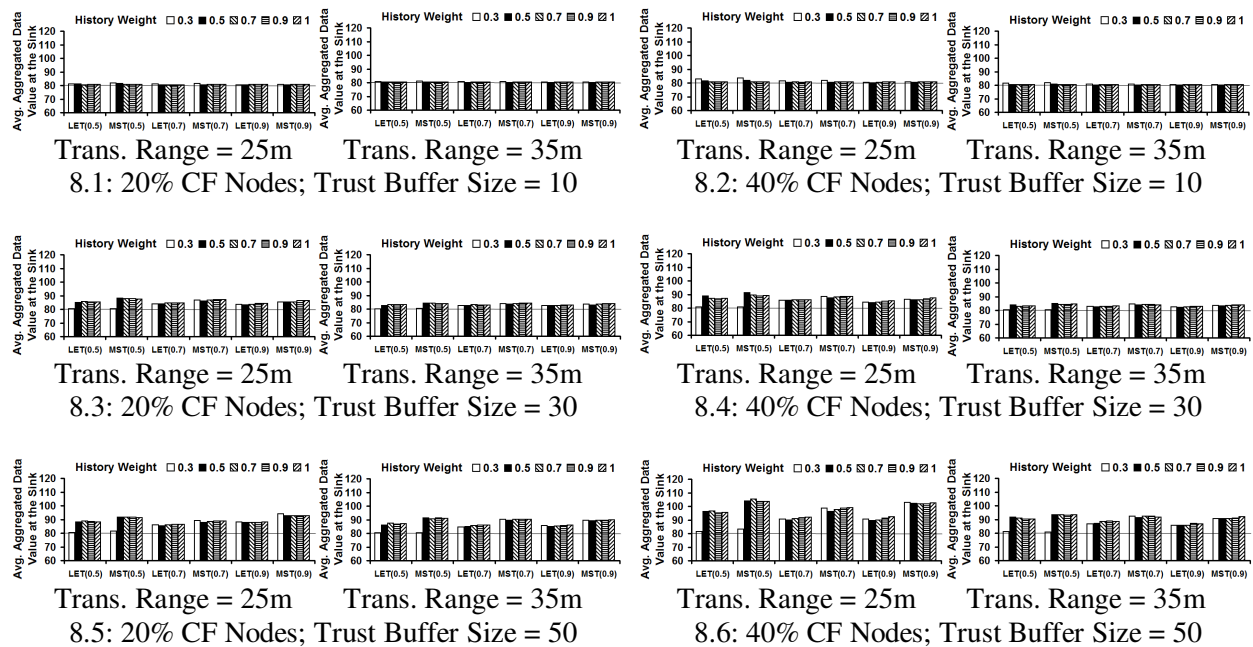

**Figure 8:** Average Aggregated Data Value at the Sink ($v_{max}$ = 3 m/s, Beacon Window Size = 10)

For both type of DG-trees, it is imperative that we operate at low Trust Buffer Size in the presence of node mobility. Given the negative influence of TSB Size on the median # rounds and data corruption, for a given TSB Size, it is more logical to operate the network at larger values of *TrustThreshold* in high node mobility conditions so that the compromised or faulty nodes can be swiftly identified when the estimated average trust score falls below the much larger *TrustThreshold* values. Also, at larger values of the Maximum TSB Size, we are more likely to end up with corrupt data because the estimated average trust score is begun to be computed only when the TSB Size reaches at least half the maximum buffer size.

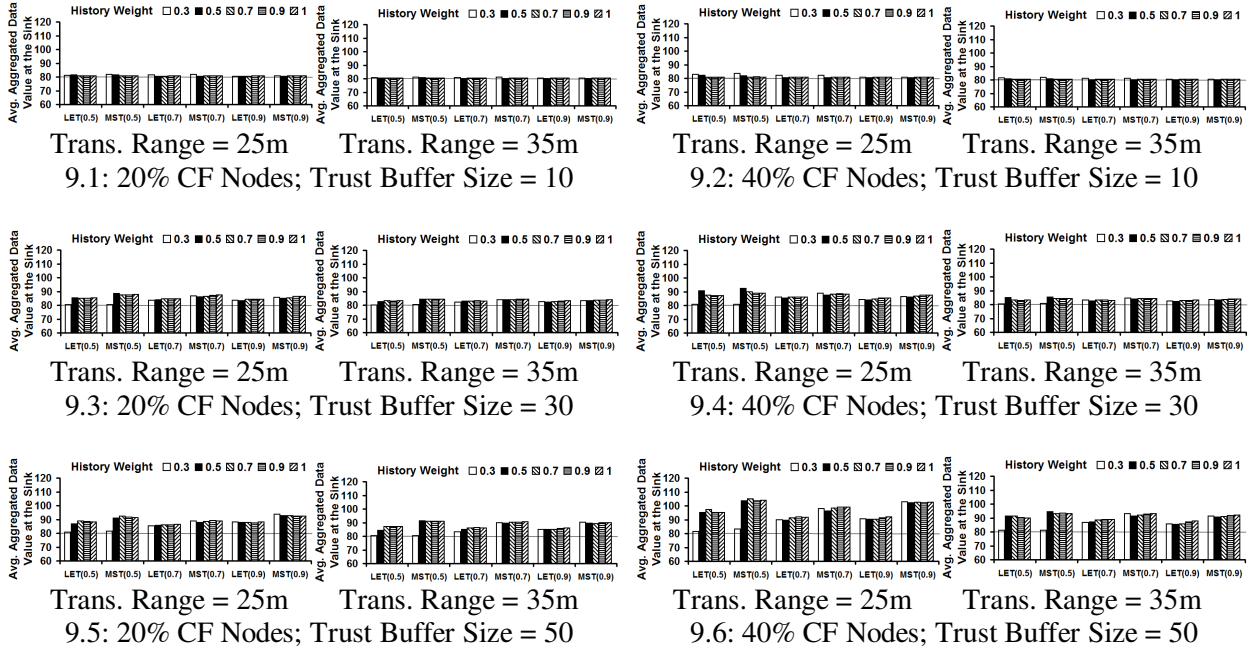

**Figure 9:** Average Aggregated Data Value at the Sink ($v_{max}$ = 3 m/s, Beacon Window Size = 50)

Note that there is no distinction among nodes (whether or not CF nodes) at the time of formation of the DG-trees. Only at the time of data aggregation, we take into consideration whether the data (individual data in case of leaf node or aggregated data in case of intermediate node) is received from a CF node and accordingly a decision to consider or discard the data is taken. At larger values of the transmission range (35m), we observe the connectivity of the network to be more likely maintained among the nodes (CF node to CF node, CF node to non-CF node, non-CF node to non-CF node) and hence we have good chances of maintaining a relatively more accurate Trust Buffer at every parent node for each of its immediate downstream child nodes. This translates to an earlier detection of the presence of CF nodes, even for larger TSB Sizes (that require a longer wait time to detect the presence of CF nodes). On the other hand, with lower transmission range of 25m, operating the network at larger *TrustThreshold* as well as a larger TSB Size, leads to more data corruption. Thus, the median # rounds to detect CF nodes is lower for transmission ranges of 35m, compared to that incurred with 25m. In the case of 25m transmission range, we observe relatively less network connectivity in the presence of CF nodes as well as node mobility. However, note that when energy loss due to transmissions and receptions are considered, the larger the transmission range, the larger is the energy loss. Hence, we could cite this as an energy-trust tradeoff.

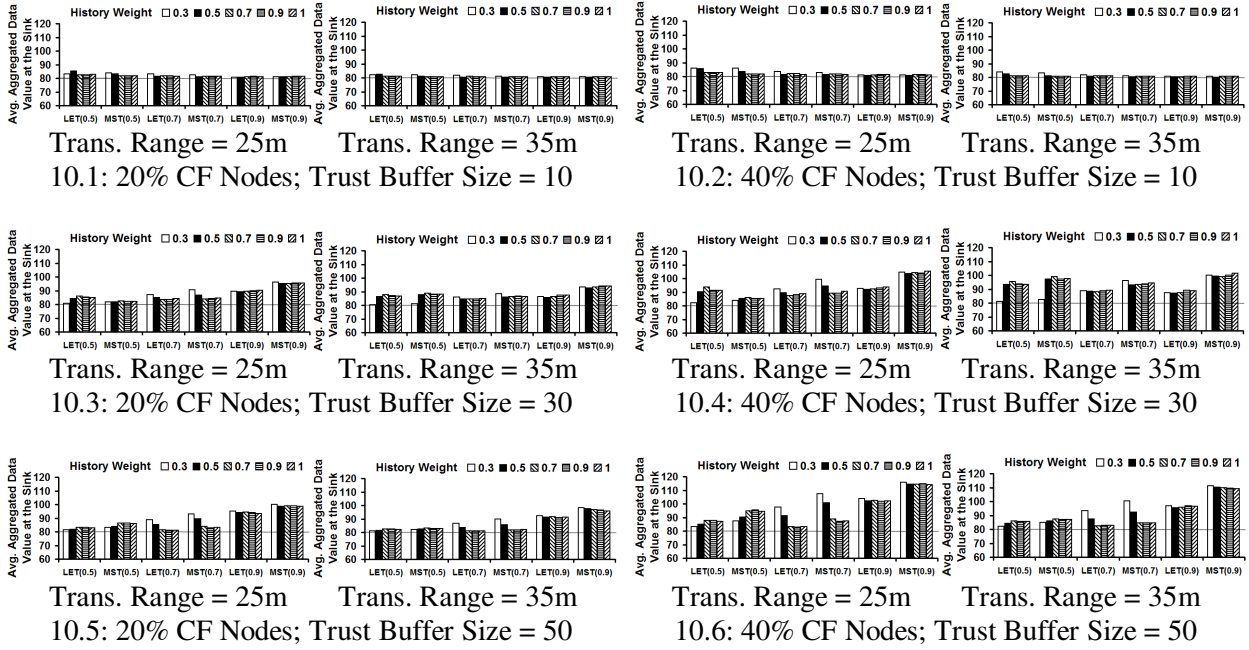

**Figure 10:** Average Aggregated Data Value at the Sink ($v_{max}$ = 10 m/s, Beacon Window Size = 10)

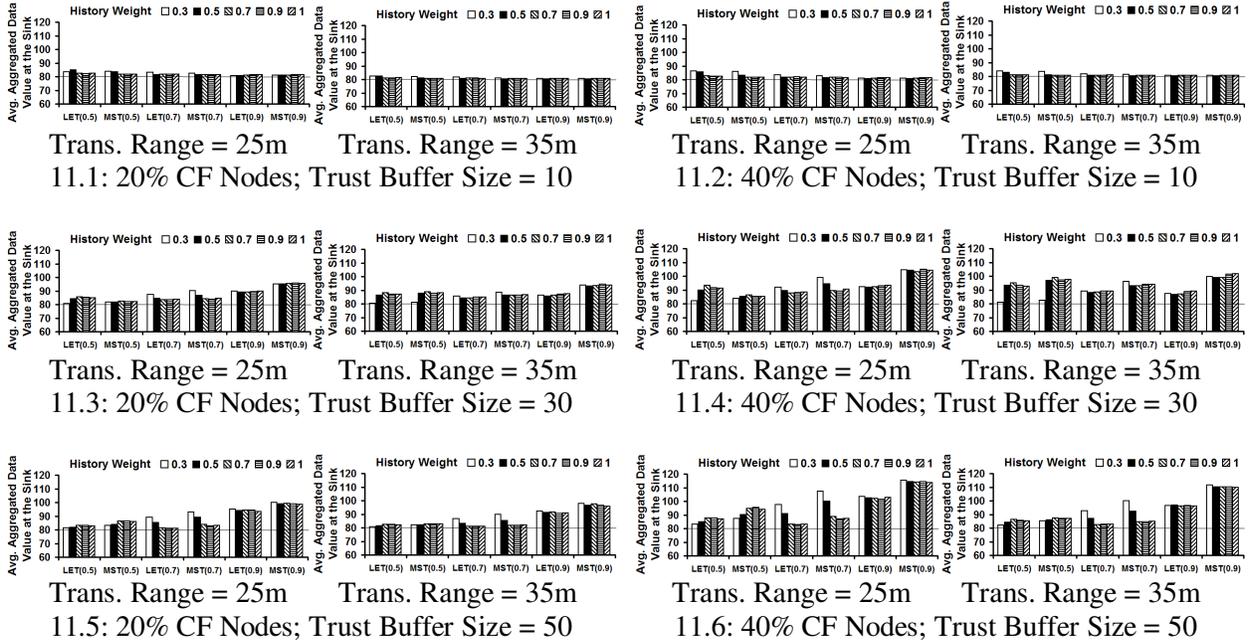

**Figure 11:** Average Aggregated Data Value at the Sink ($v_{max}$ = 10 m/s, Beacon Window Size = 50)

The larger the number of CF nodes, we would naturally expect to incur larger value for the median # rounds to detect the CF nodes. This is the case observed with TSB Size values of 10 and 30. However, for larger values of the TSB Size (50), we observe to sometimes even incur a lower value for the median # rounds, especially when operated with 40% of CF nodes (compared to 20% of CF nodes). This could be attributed to the observation that with a larger TSB Size, we could effectively detect the presence of a bunch of CF nodes in a round of data aggregation as well as discard data from the entire sub tree rooted at a CF node (the sub tree could in turn have one or more CF nodes). As a result, for larger TSB Sizes, it is possible that the CF nodes are identified as a bunch and data received from them (either in individual or aggregated form) is discarded. However, the tradeoff is that until the estimated average trust score calculations are begun, the presence of more CF nodes (especially in the case of MST-DG trees at high node mobility conditions, as the parent-child node associations are short-lived) could contribute to a relatively larger corruption of the data.

**5.8 Significant Observations from the Simulation Results**

We observe the stability-oriented link expiration time (LET)-driven spanning tree-based data gathering trees to be more suitable to assess the trust levels of the nodes and aggregate data using communication topology that bypasses (discards data) the compromised or faulty nodes. The median # rounds to detect CF nodes (and likewise the corruption in the aggregated data) incurred for the MST-based DG trees is much larger than that incurred for the LET-based DG trees, especially in high node mobility scenarios.

The Beacon Window Size used to assess the trust levels of the nodes is (by itself) not a significant parameter to influence the calculations of the estimated average weighted trust score; as long as we have a reasonable value for the Beacon Window Size (values of 10 and 50 are used in the simulations), we will be able to effectively assess the trust levels of the nodes. In the presence of moderate to high node mobility, even operating with a BW of size 10 would be more than sufficient to effectively assess the trust level of neighbor nodes.

The Trust Buffer Size (the buffer that stores the raw trust scores estimated for the nodes at subsequent rounds of the current association and previous associations, if any exist, for data gathering) plays a significant role. Larger Trust Buffer Size could contribute to collectively detecting one or more CF nodes in a single round itself (an intermediate CF node and its sub tree containing one or more CF nodes), leading to an overall reduced value for the median # rounds to detect CF nodes; but the data aggregated at the sink is also more likely to be corrupted (at least until the TB Size reaches half of its maximum size for the *Check for CF Status* to begin).

Though operating at a larger transmission range results in more energy consumption and pre-mature node failures (in energy-constrained scenarios), we observe larger transmission ranges to be effective to maintain the connectivity of the data gathering trees (especially in the presence of a larger percentage of CF nodes and high mobility scenarios), leading to lower time to detect CF nodes as well as aggregate relatively less corrupted data at the sink.

Though it is essential to recognize the trust assessment data collected during the previous associations of two nodes (a parent node and its child node), relatively more weight has to be given to the trust data collected during the current association in a DG-tree for effective trust assessment. In our simulations, we observe 70% weight to current association and 30% weight to history to be more appropriate.

The *TrustThreshold* value (below which a node is classified as CF node) has to be significantly larger (0.7-0.9) to quickly detect the CF nodes, especially in highly mobile scenarios. Lower *TrustThreshold* values contribute to much higher values for the median # rounds to detect CF nodes, especially when operated with larger TSB Size. Hence, it is more sensible to operate with a larger *TrustThreshold* value and lower TSB Size in dynamically changing communication topologies.

**6 Conclusions**

The proposed secure data aggregation (SDA) framework will be the first such comprehensive framework for defense against both insider attacks (trust evaluation model) and outsider attacks (pair-wise key establishment mechanism) for mobile sensor networks. The SDA framework provides the capability for WMSNs to develop autonomic and innate defense (self-discovery and assessment) capabilities to detect adversarial actions as well as detect and be resistant to network disruption attacks and energy depletion attacks (e.g., denial of service, false packet injection attacks). The proposed pair-wise key establishment mechanism uses the DA-tree itself as the underlying communication topology to establish pair-wise keys and does not require the need for generating a key ring for each node (pre-deployment) and path-based key establishment (post-deployment). With node mobility, the DA-trees change with time, and we anticipate the SDA framework to establish as many pair-wise secret keys as possible between the sensor nodes just based on their association in a DA-tree. Stringent mechanisms are incorporated in the design of the pair-wise key establishment mechanism to ensure confidentiality, integrity and authentication of data during aggregation. Similarly, the trust evaluation model is designed to swiftly identify the CF nodes and bypass the data aggregation process around these nodes. Simulation studies indicate that stability-based DA-trees and data gathered as part of the most recent associations between two nodes are to be preferred for effectively assessing the trust levels of the sensor nodes. With regards to the various operating parameters of the trust evaluation model, we observe the Trust Buffer Size, Trust Threshold and transmission range of the sensor nodes to play significant roles (whereas little impact is observed by varying the values of the Beacon Window Size).

## 7 Future Work

As part of future work, we will design the structure of the messages (incorporating all the fields envisioned in the design proposed earlier) exchanged as part of the pair-wise key establishment mechanism for the SDA framework and implement it in the environment of mobile sensor networks, including the energy consumption calculations. We will evaluate the energy consumption overhead associated due to the exchanges of messages as part of the pair-wise key establishment mechanism. We will also consider the impact of node mobility on the establishment of pair-wise keys between any two nodes in the network. While node mobility could increase the number of node pairs between which secret keys are newly established, it could also increase the energy consumption overhead due to exchange of messages to discover data gathering trees as well as establish/renew secret keys.

Note that the underlying assumption behind the construction of the DA-tree is that it is constructed in a secure fashion and does not comprise of compromised nodes that could launch routing attacks to attract traffic (like in warmhole attacks or sinkhole attacks) or generate traffic with multiple identities (like in Sybil attacks). The proposed trust evaluation model cannot handle nodes that could launch the above routing attacks as the model assumes that the DA-tree is securely constructed without involving nodes launching these attacks. We have illustrated the use of our recently proposed protocols to construct DA-trees in a distributed fashion. Developing a protocol to construct DA-trees that are robust to routing attacks (like warmhole, sinkhole or sybil attacks) in mobile sensor networks is beyond the scope of this paper. We have not come across any such protocol so far for mobile sensor networks. As part of future work, we will develop secure data aggregation protocols that are robust to these routing attacks during the process of constructing the DA-tree itself, and evaluate the performance of the trust evaluation model in the presence of compromised nodes that launch the routing attacks.